\documentstyle[aps,multicol,epsfig]{revtex}

\begin{document}
\title{Nonequilibrium orientational patterns in two-component Langmuir 
monolayers}
\author{Ramon Reigada$^{1},$ Alexander S. Mikhailov$^{2},$ and Francesc 
Sagu\'es$%
^{1} $}
\address{$^{1}$Departament de Qu\'{\i}mica-F\'{\i}sica, Universitat de 
Barcelona,\\
Avda. Diagonal 647, 08028 Barcelona, Spain\\
$^{2}$Abteilung Physikalische Chemie, Fritz-Haber-Institut der\\
Max-Planck-Gesellschaft,\\
Faradayweg 4-6, 14195 Berlin, Germany}
\maketitle

\begin{abstract}
A model of a phase-separating two-component Langmuir monolayer in the
presence of a photo-induced reaction interconvering two components is
formulated. An interplay between phase separation, orientational ordering
and treaction is found to lead to a variety of nonequilibrium
self-organized patterns, both stationary and traveling. Examples of the
patterns, observed in numerical simulations, include flowing droplets,
traveling stripes, wave sources and vortex defects. 

\vspace{8pt} PACS: 82.20.Wt, 47.54.+r, 82.45.Mp, 68.47.Pe \vspace{8pt}
\end{abstract}


\vskip1cm



\section{Introduction}

\label{intro}

Spurred by experimental and technological developments in supramolecular
chemistry and biophysics, there is a raising interest to study
nonequilibrium structures related to self-organization phenomena in weakly
condensed systems. In such systems, attractive potential interactions
between constituent molecules are essential. Equilibrium structures in soft
matter correspond to minima of free energy, and are caused by the
competition between short-range attractive and long-range repulsive
interactions \cite{and1}. In contrast to this, nonequilibrium patterns
require permanent supply of energy and correspond to persistent (stationary
or time-dependent) kinetic states of a system\cite{mikhsc}. Typically, they
result from an interplay between reactions, diffusion and phase transitions.
Examples of nonequilibrium pattern formation in reactive soft matter include
stationary Turing-like patterns in phase-separating binary polymer mixtures
with chemical reactions \cite{glo2,moto,tcong1} and in monomolecular
adsorbates on metal surfaces \cite{dewell,mexPRE}. Theoretical analysis for
two-component reactive adsorbates \cite{hild,mikhstat} and for three-component
reactive polymer systems \cite{okuz,ohta2} has further shown that not only
stationary patterns, but also traveling and standing waves are possible.

Langmuir films are monomolecular layers of amphiphilic molecules on an
air-water interface. Such films are a classical example of soft matter and,
generally, it should be expected that, in the presence of chemical reactions
and energy flows, they would form nonequilibrium patterns.
Though equilibrium properties of Langmuir monolayers are thoroughly
investigated (see \cite{kag1}), nonequilibrium phenomena in these systems
still remain less explored. In an interesting series of experiments, Tabe
and Yokoyama have studied illuminated Langmuir monolayers of amphiphilic
derivatives of azobenzene by means of the Brewster-angle microscopy \cite
{tabe1,tabe2a,tabe2b,tabe3}. In these experiments, transitions between {\it trans}
and {\it cis} conformations of individual molecules were photo-induced by
polarized light of a selected wavelength. Since physical properties of the
two conformations are different, the {\it trans }and {\it cis }isomers
essentially represented two different species. In addition to photo-induced
periodic stationary patterns, these experiments have shown, for the first
time, spontaneously emerging patterns of propagating waves of molecular
reorientation under appropriate illumination conditions.

In the Letter \cite{our}, we proposed a model of reactive two-component
Langmuir monolayers with orientational ordering. This model was
taking into account phase separation in the two-component system,
reaction interconverting both species, diffusion of reactants and processes
of orientational ordering. Interactions between the components of the
monolayer resulted not only from the positional, but also from the
orientational order of the hydrophobic tails of constituent molecules,
determined by their tilt. We have shown that this model
already reproduces nonequilibrium traveling-wave patterns which arise as a
consequence of a Hopf bifurcation with a finite wavenumber. To simplify the
analysis, it was assumed that the azimuthal orientation of molecules
remained fixed and uniform, so that only their tilts could vary. Moreover,
some orientational-order contributions to the free energy of the
monolayer were neglected. The inclusion of azimuthal variations is, however,
important for a comparison with the experimental data yielded by the
Brewster-angle microscopy that is sensitive to the local azimuthal orientational
ordering.

The aim of the present paper is to formulate and to study a more general
model that contains both orientational variables, and includes bend and
splay distortion terms. Our analytical and numerical investigations show
that this model has a significantly different phase diagram and new kinds of
nonequilibrium patterns are possible here.
After introduction of the model in Sec. \ref{model}, we investigate in 
Sec. \ref{noreaction} the behaviour of the system in the equilibrium case, when
the reaction is absent. The bifurcation analysis of the uniform steady state
of the nonequilibrium system under illumination is presented in Sec. \ref{linear}.
Numerical simulations, revealing the formation of such
nonequilibrium patterns as traveling droplets and stripes, stationary
splay-defects, and complex azimuth reorientation kinetics, are reported in
Sec. \ref{reaction}. The paper ends with the conclusion and the discussion
of the obtained results.

\section{The Model}
\label{model}

\begin{figure}[tbh]
\vspace{-0.2in}
\begin{center}
\epsfxsize = 3.in
\epsffile{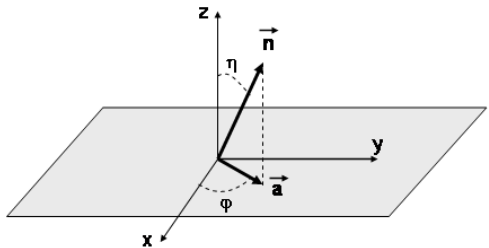}
\end{center}
\vspace{-0.2in}
\caption{Schematic illustration of the variables used to describe the tilted
elongated molecules in the monolayer.}
\label{figmodel}
\end{figure}

We consider a monolayer formed by two diffusive immiscible components, $A$
and $B$, that have strongly different shapes. Modeling the situation in the
photo-isomerization experiments, molecules $A$ are supposed to have an
elongated shape with a long tail (the {\it trans} isomer) and molecules $B$
to have a crumpled conformation (the {\it cis} isomer). According to this
assumption, only molecules $A$ are subject to orientational order, whereas
molecules $B$ play essentially the role of passive dilution with respect to
such ordering. Furthermore, a photo-induced reaction interconverting $A$ and 
$B$ molecules is considered. The total concentration of components $A$ and 
$B$ in the monolayer is assumed constant. Therefore, the local composition
of the monolayer is characterized by the concentration order parameter $c$
representing the local fraction of molecules $A$ (so that $1-c$ gives the
local fraction of molecules $B$). The local orientational order is described
by the order parameter $\vec{a}$, that corresponds to the projection of the
local mesoscopic average of the unit molecular director $\vec{n}$ of the
elongated molecules $A$ onto the monolayer plane (see Fig. \ref{figmodel}).
The vector $\vec{a}$ is defined by its modulus $a=|\vec{a}|=\sin \eta$,
where $\eta$ is the tilt angle, and its azimuth angle $\varphi$. 

After we have introduced the two order parameters $c$ and $\vec{a}$ needed
to describe the system, the kinetic equations for their evolution should be
formulated. Following the mesoscopic approach \cite{our}, the evolution
equation for $c$ reads as 
\begin{equation}
\frac{\partial c}{\partial t}=D\nabla ^{2}c+\frac{D}{k_{B}T}\nabla \left[
c(1-c)\nabla \mu \right] +{\cal R}(c).
\label{kinc}
\end{equation}
Here $D$ is the diffusion coefficient, $T$ is temperature, and $\mu $ is the
chemical potential defined as $\mu =\delta {\cal F}/\delta c$, where ${\cal F}$
is the free energy functional that will be specified below. In the
presence of nonpolarized light, the reactive term ${\cal R}(c)$ is given by 
\cite{laidler,maack2,maack}, 
\begin{equation}
{\cal R}(c)=\left[ Ig(\lambda )+k_{-1}^{0}\right] (1-c)-\left[ If(\lambda
)+k_{1}^{0}\right] c,
\label{kakb}
\end{equation}
where the first term corresponds to the conversion of molecules $B$ into
molecules $A$ and the second term describes the reverse conversion process.
In this equation, $I$ is the light intensity, $\lambda $ is the wavelength of
light, and $f(\lambda )$ and $g(\lambda )$ are related to the surface molar
extinction coefficients and quantum yields for the conformations of $A$ and 
$B$, respectively. For moderate and high illumination intensities,
photo-induced conversion dominates over the thermal conversion
processes, so that the thermal rate constants $k_{\pm 1}^{0}$ can be
neglected. In this case, the ratio of conversion rates for forward and
backward processes is independent of temperature or light intensity, and
determined only by the light wavelength $\lambda $. For simplicity, we shall
assume in our subsequent analysis that the wavelength $\lambda $ is chosen
in such a way that $f(\lambda )=g(\lambda )$ . Under this condition,
equation (\ref{kakb}) takes the form 
\begin{equation}
{\cal R}(c)=k(1-2c),
\label{kakb2}
\end{equation}
where $k=If(\lambda )=Ig(\lambda )$ is the reaction constant proportional to
the intensity of the applied light.

The kinetic equation for the local orientation vector $\vec{a}$ is chosen as 
\begin{equation}
\frac{\partial \vec{a}}{\partial t}=-\Gamma \frac{\delta {\cal F}}{\delta 
\vec{a}}-kq(c)\vec{a}.
\label{kina}
\end{equation}
The first term on the right side corresponds to relaxation with a typical
relaxation time $\Gamma ^{-1}$. The second term takes into account that the
reaction, interconverting one molecular conformation to another, also
affects the local average orientation of molecules.

The choice of the function $q(c)$ should be based on the kinetic analysis
with respect to the orientation of molecules involved in the reactive
process. In our model, the reaction $B\longrightarrow A$, transforming
crumpled molecules into the elongated form, is assumed to be strongly
energetically activated by light. This means that new molecules $A$ would be
created with high initial energy and can adopt any orientation $\vec{a}$,
independent of the orientation of surrounding molecules. Therefore, the
orientation of newly created molecules $A$ is random and the statistical
average of $\vec{a}$ is zero. Because any conversion from an elongated
molecule $A$ to the crumpled molecule $B$ decreases the local order and the
reverse conversion process does not increase it, the overall reaction effect
is that it tends to destroy orientational ordering.
Under such assumptions, local evolution for the orientation momentum 
$c\vec{a}$ is described by a simple equation $\partial _{t}(c\vec{a})=-kc\vec{a}$,
that corresponds to the loss of $c\vec{a}$ when molecules $A$ transform into 
$B$, having no defined orientation. Splitting $\partial _{t}(c\vec{a})$ and
substituting the local variation of $c$ due to reaction, $\partial
_{t}c=k(1-2c)$, leads to $\partial _{t}\vec{a}=-k\vec{a}(1-c)/c$. Therefore,
we have $q(c)=(1-c)/c$. This function $q(c)$ is used below.

The energy functional ${\cal F}$ is constructed in terms of the order
parameters $c$ and $\vec{a}$ of the system. We decompose the energy
functional into two parts: one that accounts for the affinity between the
two isomers in the monolayer, and the other due to the tail-tail
orientational interactions. The first of these contributions,
${\cal F}_{c}$,
depends exclusively on $c$, whereas the tail-tail interaction,
${\cal F}_{or}$, is sensitive to both $c$ and the orientation of the elongated
molecules. Accordingly, ${\cal F}={\cal F}_{c}(c)+{\cal F}_{or}(c,\vec{a})$.
Note that entropic contributions are not considered, since they have been
directly included as the diffusive term in the kinetic equation (\ref{kinc}).
The expression for ${\cal F}_{c}$ reads as \cite{our}, 
\begin{equation}
{\cal F}_{c}=\int {dxdy}\left[ -\tilde{\chi _{0}}c^{2}+\frac{1}{2}
\tilde{\chi _{2}}\left( \nabla c\right) ^{2}\right] ,
\label{f8}
\end{equation}
and describes phase separation due to lateral interactions between
molecules. In the limit of short-range lateral interaction potentials, $%
\tilde{\chi _{0}}$ is determined by its strength, and $\tilde{\chi _{2}}$ can be
estimated as $\tilde{\chi _{2}}\approx \frac{1}{2}\tilde{\chi 
_{0}}r_{0}^{2}$, where $r_{0}$ is the characteristic
radius of the interaction \cite{mikh6}.
Near the critical point of the equilibrium phase separation, the
combination of Eqs. (\ref{kinc}) and (\ref{f8}) is equivalent to the usual
Cahn-Hilliard equation with the Landau free energy functional \cite{mexPRE}.

The part of the free energy functional associated with the distortion of the
orientation of tilted molecules in the monolayer can be written as 
\begin{equation}
{\cal F}_{or} =\int {dxdy}\left[-
\frac{1}{2}\tilde{p}(c)a^{2}+\frac{1}{4}\tilde{\beta}a^{4}+ 
\frac{\tilde{K_{s}}}{2}\left( \vec{\nabla}\cdot \vec{a}\right) 
^{2}+\frac{\tilde{K_{b}}}{2}\left( \vec{\nabla}\times \vec{a}\right) 
^{2}+\tilde{\Lambda}c\left( \vec{\nabla}\cdot \vec{a}\right)\right].
\label{for}
\end{equation}
The first two terms correspond to the Landau expansion up to the quartic
term for the modulus of the orientation vector \cite{kag1}. This expansion
is only justified for sufficiently small $\eta $, since in this case $a=\sin
\eta \approx \eta $ is small as well. Generally, all coefficients in the
Landau decomposition should depend on the local concentration $c$. We shall,
however, consider only weakly nonuniform states, where local deviations of
the concentration $c$ from the uniform stationary state $\overline{c}=1/2$
are small. Therefore, we neglect the dependence of the coefficient
$\tilde{\beta}$ on the variable $c$. However, the coefficient $\tilde{p}$ of the
quadratic term in the Landau free energy is already small near the
instability and its dependence on $c$ must be retained. For condensed
phases, lowering the lateral pressure of a Langmuir monolayer leads to an
increase of its equilibrium tilt \cite{kag1}. Since we assumed that
molecules $B$ play a role of passive dilution for the tilted molecules $A$,
decreasing $c$ is roughly equivalent to decreasing the lateral pressure.
Thus, we choose a linear dependence $\tilde{p}(c)=\tilde{\pi }_{0}+2\tilde{%
\alpha}(0.5-c)$, where $\tilde{\alpha}$ is a positive coefficient, and $%
\tilde{\pi}_{0}$ is a decreasing function of the lateral pressure which
determines the equilibrium tilt in the monolayer with $c=\overline{c}$.

The other contributions in Eq. (\ref{for}) stand for the bend and splay
distortion terms \cite{meyer}. More specifically, the third and fourth terms
correspond to the classical Frank elasticity terms, that account for the
splay and bend distortions, respectively. Normally, one takes the single
Frank constant approximation ($\tilde{K}=\tilde{K_{s}}=\tilde{K_{b}}$). The
fifth term is the lowest-order splay contribution, $(\vec{\nabla}\cdot 
\vec{a})$, that appears coupled to the composition order parameter $c$.
Although
some authors \cite{pets,najjar} prefer to couple the linear splay term to
the tilt angle, we follow the suggestion by Selinger {\it et al.}\cite
{selinger} for a two-component monolayer undergoing phase separation, as it
is in our case. A similar approach was taken by Tabe {\it et al.} \cite
{tabe2a,tabe2b} who coupled the linear splay term with a certain order parameter
related to the molecular density that varies across the monolayer. In
general, for sufficiently strong coupling $\tilde{\Lambda}$, the linear
splay term destabilizes uniformly oriented phases, leading to equilibrium
nonhomogeneous splayed states (see Sec. \ref{noreaction}). Finally, notice
that a term $(\vec{\nabla}\times \vec{a})$ linear with respect to bend
distortion is not considered, because it is not permitted by symmetry in
nonchiral Langmuir monolayers.

Summarizing, the model presented here can be viewed as a Cahn-Hilliard
equation for the composition variable $c$, coupled to a relaxational
equation for the orientation order parameter $\vec{a}$. The contributions to
the free energy have been derived considering that $c$ is close to its
stationary uniform solution and for sufficiently small tilt angles $\eta$.

The analysis can be simplified by appropriately adimensionalizing energy,
time and space. Energy is measured in units of $k_{B}T$, time in units of
the relaxational time $(\Gamma k_{B}T)^{-1}$, and spatial coordinates are
rescaled with the relaxational length $\sqrt{D/(\Gamma k_{B}T)}$. The model
is then characterized by the dimensionless parameters $\chi _{0}=\tilde{\chi
_{0}}/k_{B}T$, $\chi _{2}=\tilde{\chi _{2}}\Gamma /D$, $\pi _{0}=\tilde{\pi 
}%
_{0}/k_{B}T$, $\alpha =\tilde{\alpha}/k_{B}T$, $\beta =\tilde{\beta}/
k_{B}T$%
, $\kappa =k(\Gamma k_{B}T)^{-1}$, $K=\tilde{K}\Gamma /D$ and $\Lambda =%
\tilde{\Lambda}\sqrt{\Gamma (Dk_{B}T)^{-1}}$. With this choice, the final
equations for the evolution of $c$ and the two components of vector 
$\vec{a}$
read as
\begin{eqnarray}
\frac{\partial c}{\partial t} &=&\nabla ^{2}c+\nabla \left[ c(1-c)\nabla \mu
\right] +\kappa (1-2c),  \nonumber \\
\frac{\partial \vec{a}}{\partial t} &=&p(c)\vec{a}-\beta a^{2}\vec{a}%
+K\nabla ^{2}\vec{a}+\Lambda \vec{\nabla}c-\kappa \frac{1-c}{c}\vec{a},
\label{kinfull}
\end{eqnarray}
where
\[
\mu =-2\chi _{0}c-\chi _{2}\nabla ^{2}c+\alpha a^{2}+\Lambda \vec{\nabla}
\cdot \vec{a},
\]
and $p(c)=\tilde{p}(c)(k_{B}T)^{-1}=\pi _{0}+2\alpha (0.5-c)$. This is the
mathematical model which will be investigated below.

In order to obtain numerical results from the model,
we numerically integrate Eqs. (\ref{kinfull}) on a $100\times 100$
square grid, using an explicit Euler scheme with a mesh size $\Delta x$ and
a time step $\Delta t$ that assure a good numerical accuracy. Periodic
boundary conditions are chosen to model the behaviour in a large system far
from the boundaries. As initial conditions, small random perturbations
around the stationary uniform states of the system $\overline{c}$ and
$\overline{a}$, and a random distribution of azimuth angles $\varphi \in
(0,2\pi )$ are taken.
To display simulation results, snapshots of the patterns after the
transients are given in the figures. Each figure consists of two panels: the
left panel shows in grey scale the spatial distribution of the composition
variable $c$, with larger values corresponding to the darker color, and the
right panel is used for visualization of the orientational field $\vec{a}$.
The local directions of this field are visualized by using small arrows.
Note that, for technical reasons, such arrows could not be used to indicate
the states of all grid points in the simulations and therefore the
visualization of the azimuthal orientation is rough. The grey color in the
right panels is used to display the local tilt $a$, and again, darker regions
correspond to the larger values of this variable. In some cases, videos of
time-dependent patterns are additionally provided.

\section{Equilibrium Patterns}
\label{noreaction}

Before addressing the nonequilibrium cases, we show some
examples of equilibrium pattern formation in the considered system. The
equilibrium conditions correspond to absence of illumination and are
realized if all reactive terms in Eqs. (\ref{kinfull}) are omitted. One of
the limitations of our previous simpler model \cite{our} was that it did not
exhibit any equilibrium pattern formation, despite the experimental evidence
of spontaneous generation of striped patterns in non-illuminated monolayers 
\cite{tabe1,tabe2a,tabe2b}. The linear splay term included in Eq. (\ref{for}) can
already lead to the formation of equilibrium orientational structures.

\begin{figure}[htb]
\vspace{-0.05in}
\begin{center}
\epsfxsize = 3.4in
\epsffile{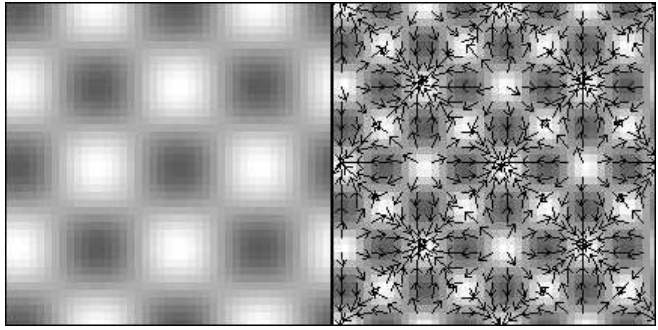}
\end{center}
\caption{Concentration (left panel) and orientation (right panel)
fields in the equilibrium pattern representing an array of splay defects for
$\Lambda=2$, $K=0.25$, $\pi_0=\alpha=0$, $\beta=4$, $\chi_2=1$, and $\chi_0=1$
in absence of reaction ($\kappa=0$). The grey color is used in the right
panel to display the local tilt $a$ of the molecules.
In both panels, darker regions correspond to higher values of the displayed variables.}
\label{figselingera}
\end{figure}

\begin{figure}[htb]
\vspace{-0.05in}
\begin{center}
\epsfxsize = 3.4in
\epsffile{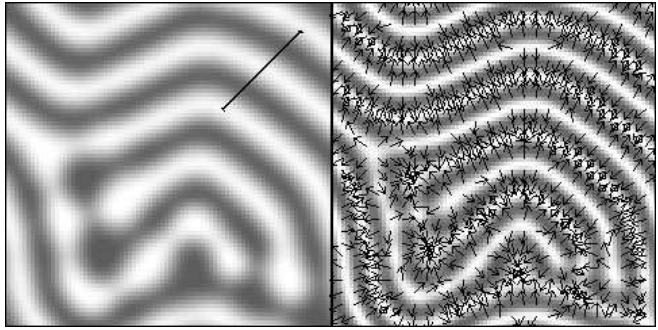}
\end{center}
\caption{Equilibrium stripe pattern for $\Lambda=2$, $K=0.25$,
$\pi_0=\alpha=0$, $\beta=4$, $\chi_2=1$, and $\chi_0=2$, in absence of reaction ($\kappa=0$).}
\label{figselingerb}
\end{figure}

\begin{figure}[htb]
\vspace{-0.1in}
\begin{center}
\epsfxsize = 2.4in
\epsffile{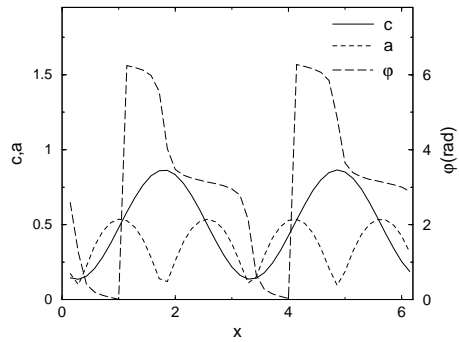}
\end{center}
\caption{Profiles of composition, tilt and azimuth angle along
the cross section of the stripe pattern indicated by a line segment in the left panel
of Fig. \ref{figselingerb}.
The azimuthal angle is measured with respect to the positive direction of the axis $x$.}
\label{figperfstripes}
\vspace{-0.1in}
\end{figure}

\begin{figure}[htb]
\vspace{-0.1in}
\begin{center}
\epsfxsize = 2.4in
\epsffile{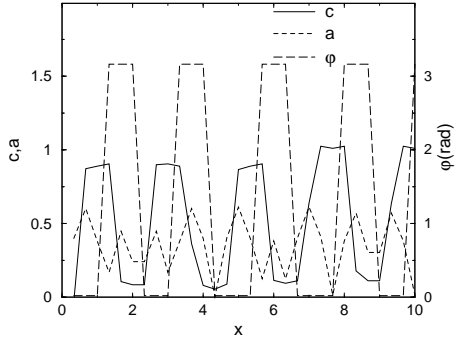}
\end{center}
\caption{Profiles of composition, tilt and azimuth angle along a cross
section of the equilibrium soliton-like pattern with sharp wall domains
found for $\chi_0=8$ and the same other parameters as in Figs. \ref{figselingera}
and \ref{figselingerb}.}
\label{figperfsoliton}
\vspace{-0.1in}
\end{figure}

In the limit of $\alpha \rightarrow 0$ (i.e., in the absence of the
quadratic Landau term for the tilt variation), the nonreactive version of
the present model is similar to the description proposed by Selinger {\it et
al.} \cite{selinger} for nonchiral Langmuir monolayers. In their study,
three different nonuniform phases (sinusoidal stripes, soliton stripes and
square lattice of vortices) were found for smectic films undergoing chiral
symmetry breaking under variation of a control parameter (corresponding to
temperature), and similar phases for multicomponent Langmuir monolayers were
predicted. As illustrated in Figs. \ref{figselingera},
\ref{figselingerb}, \ref{figperfstripes}
and \ref{figperfsoliton}, our model reproduces these regimes under
appropriate choices of the parameter $\chi_{0}$ specifying the relative
intensity of energetic lateral interactions.

When the characteristic energy of lateral interaction is much weaker than
the orientation energy, a square array of alternating ``inward'' and
``outward'' splay defects (vortices) is formed (Fig. \ref{figselingera}).
The ``inward''
defects represent regions rich in elongated molecules $A$ which are oriented
towards the center of the defect. In analogy to the ``escape to the third
dimension'' found in defects in $3D$ nematics \cite{esc3d}, the tilt in the
center of these defects is almost zero, reducing the Frank elastic energy
near that point. The ``outward'' splay defects are poor in elongated
molecules $A$, but also exhibit vanishing tilt in their centers. Increasing
the lateral interactions with respect to the splay coupling, the system
organizes into a stripe phase with smooth variations of $c$ and $a$
(see Fig. \ref{figselingerb}).
The profiles of variation of c, $a$ and $\varphi$ across a stipe are
displayed in Fig. \ref{figperfstripes}.
The equilibrium stripe patterns are similar to those
observed in the experiments in absence of illumination \cite{tabe2a,tabe2b,proper}.
The modulation of the tilt (absent in the model by Selinger {\it et al.}
because tilt variations were not included there) has a spatial frequency
twice that of the azimuthal and concentration modulations. Moreover, in
agreement with the experiments the amplitude of the tilt angle modulation
depends on the stripe size, decreasing as the stripe widens (this is
observed, for example, by decreasing the parameter $\Lambda$).

Strong lateral attractive interactions (i.e., large interaction strengths
$\chi_{0}$) lead to the formation of a striped phase with sharp wall domains
(''soliton stripes'') in the modulation of composition, tilt and azimuth,
which has also been reported by Selinger {\it et al.} \cite{selinger}. In
Fig. 5 profiles of variation of c, $a$ and $\varphi $ across such a stripe
are plotted. Note again the double frequency modulation of the tilt with
respect to the composition and azimuth variation.

We have checked that the inclusion of the quadratic Landau term for the tilt
variable, neglected in the simulations described in this section, does not
lead to significative differences in the properties of equilibrium patterns.

\section{Nonequilibrium Phenomena: Linear Stability Analysis}
\label{linear}

Full numerical exploration of the parameter space of the model is difficult
because of the large number of relevant parameters. The linear stability
analysis of the uniform stationary states can provide indications of what
regions of the parameter space are worth being examined in search for a
particular kind of a pattern. The stationary states of the system (\ref
{kinfull}) are $c=\overline{c}=1/2$ and $a=\overline{a}=\sqrt{\left( \pi
_{0}-\kappa \right) /\beta }$, provided that $\pi _{0}>\kappa $. The
azimuthal orientation is arbitrary in such a state. For convenience, we
choose $\overline{a_{x}}=\overline{a}$ and $\overline{a_{y}}=0$. The linear
stability of these uniform solutions is performed by adding small plane-wave
perturbations $\delta c$, $\delta a_{x}$ and $\delta a_{y}$ proportional to 
$\exp (iq_{x}x+iq_{y}y+\gamma (\vec{q})t)$, and linearizing 
Eqs. (\ref{kinfull}). This leads to the linearization matrix 
\begin{equation}
{\cal L}=\left( 
\begin{array}{ccc}
-\frac{q^{2}}{2}\left( -\chi _{0}+\frac{q^{2}}{2}\chi _{2}+2\right) -2\kappa
& -\frac{q^{2}}{2}\left( \alpha \overline{a}+i\frac{\Lambda }{2}q_{x}\right)
& -i\frac{\Lambda }{4}q_{y}q^{2} \\ 
2\overline{a}(2\kappa -\alpha )+iq_{x}\Lambda & \pi _{0}-\kappa -3\beta 
\overline{a}^{2}-Kq^{2} & 0 \\ 
iq_{y}\Lambda & 0 & -Kq^{2}
\end{array}
\right) , 
\label{3x3}
\end{equation}

where $\vec{q}=(q_x,q_y)$ is the wavevector of the considered mode. The first line in
the matrix corresponds to the composition variable $\delta c$ and the next
two lines stand for the orientation components $\delta a_{x}$ and $\delta
a_{y}$.

The (complex) rates $\gamma (\vec{q})$ of various modes are determined by
the eigenvalues of the linearization matrix ${\cal L}$. The unstable modes
are identified like those with Re$(\gamma (\vec{q}))>0$. If the imaginary part of
$\gamma(\vec{q})$
is not zero for the first unstable mode, we have a wave instability (a Hopf
bifurcation with a finite wavenumber), resulting in traveling or standing
waves (cf. \cite{wal}). On the other hand, if Im$(\gamma (\vec{q}))=0$ for
the first unstable mode, a Turing instability leading to nonequilibrium
stationary periodic patterns is realized. The values of $\gamma (\vec{q})$
are yielded as the roots of the characteristic equation associated with the
matrix ${\cal L}$. This characteristic equation is, however, cubic and
therefore its analytical solution is possible only in some special cases.

\begin{figure}[tbp]
\begin{center}
\par
\def\epsfsize#1#2{0.3\textwidth}
\leavevmode
\epsffile{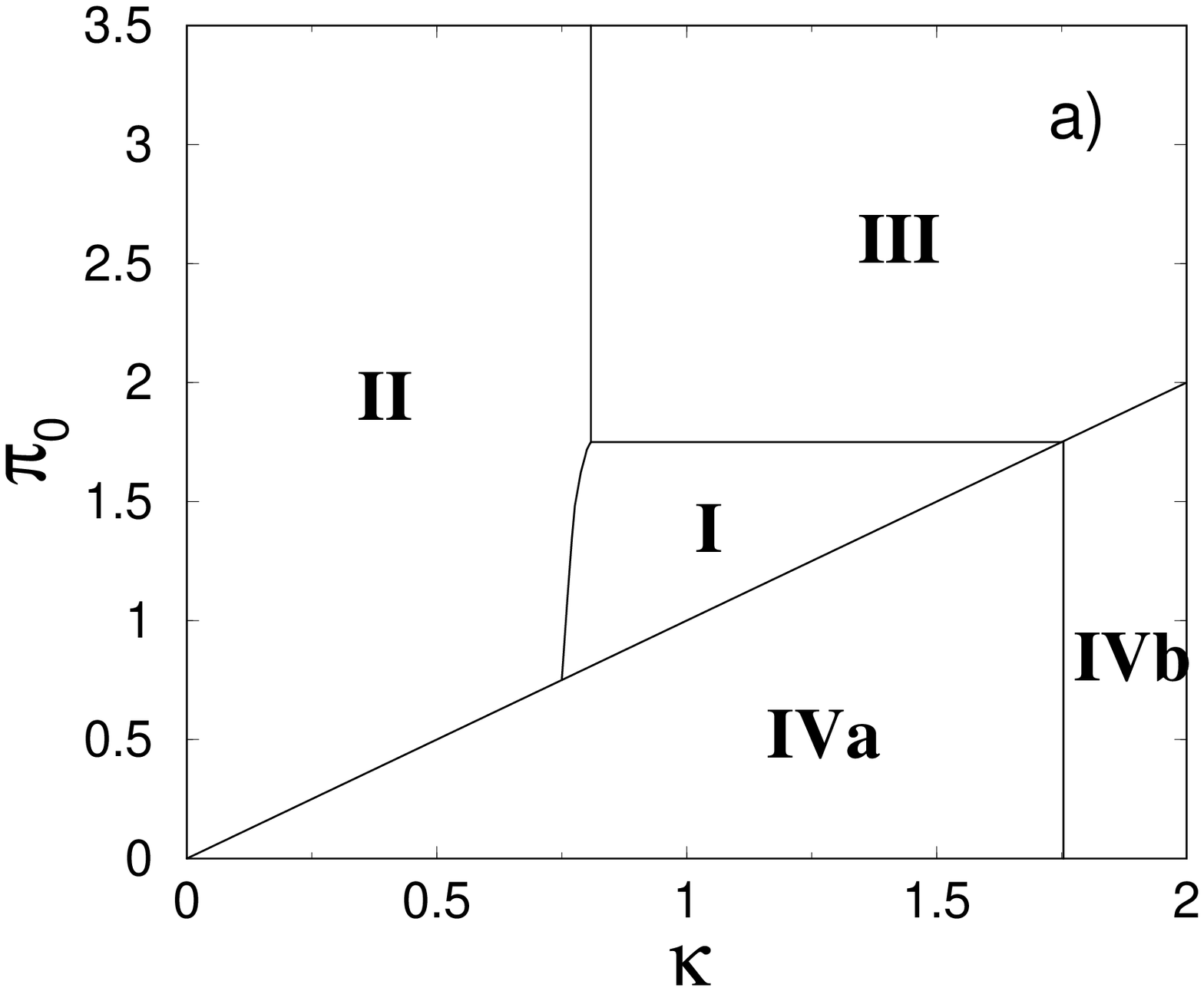}
\def\epsfsize#1#2{0.3\textwidth}
\epsffile{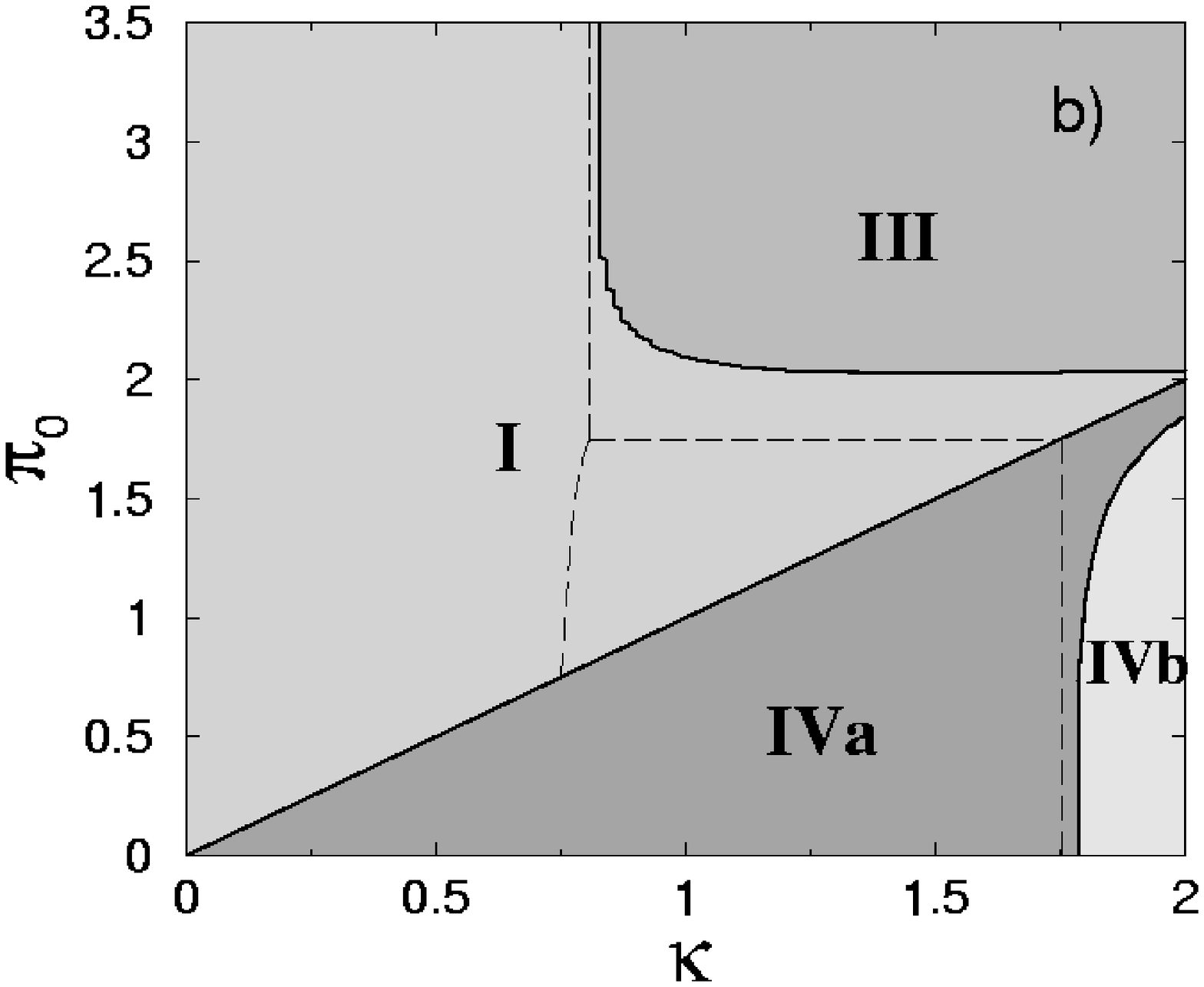}
\end{center}
\caption{Phase diagrams of the considered model for the parameter values
$\chi_{2}=0.0052$, $\chi_{0}=2.27$, $\alpha=1.5$, $\beta=2$
in the limit $K\rightarrow 0$ (a) without splay coupling and (b)
with weak splay coupling ($\Lambda=0.01$). The diagram (b) is
obtained by the linear stability analysis with respect to the
perturbations with $q_{y}=0$. Dashed lines in this diagram show
the boundaries of the respective regions in absence of splay coupling (a).
Other notations are explained in the text.}
\label{figdiagfas}
\end{figure}

If splay coupling is absent ($\Lambda=0$), the stability analysis is
simplified and becomes equivalent to that of the previously studied reduced
model \cite{our}. The phase diagram for the non-splay case in the plane
($\pi_{0}$,$\kappa$) is presented in Fig. \ref{figdiagfas}a. We see five
different regions, whose boundaries and marginal wave numbers can be
obtained analytically \cite{our}. Region $I$ corresponds to the wave
instability regime, where traveling or standing waves are expected. In
region $II$ we have the Turing instability region, where stationary
droplet-like structures with periodic variation of both local concentration
and tilt are found. Region $III$ corresponds to the uniform tilted phase.
When $\pi_{0}<\kappa $, only nontilted phases ($\overline{a}=0$) exist: a
nonuniform phase due to a Turing instability in region $IVa$ and stable
uniform phase in region $IVb$. 

In order to understand the effects of the linear splay term, that was
neglected above, we perform the stability analysis of the full model
equations for a fixed wave plane $q_{y}=0$. In this case, the variable $a_{y}$
is decoupled from the other two degrees of freedom (actually, its
dynamics is exclusively governed by the elastic damping term, see Eq. (\ref{3x3})), 
so that the stability analysis is reduced to a $2\times 2$ problem.
The following results, summarized in Fig. \ref{figdiagfas}b, are obtained:
For nontilted phases ($\pi_{0}<\kappa$), the effect of the linear splay
term is to move to larger reaction rates $\kappa$ the boundary between
the Turing-instability region ($IVa$) and the uniform nontilted phase 
($IVb$). For tilted phases ($\pi_{0}>\kappa$), the changes are more profound.
All unstable modes in this area of the parameter space have now a
nonvanishing imaginary part Im$(\gamma(\vec{q}))$. Therefore, the region $%
II $ with Turing instability and stationary tilted patterns completely
disappears and becomes replaced by the region $I$ with traveling waves.
Additionally, the region $III$ occupied by the tilted uniform phase is
reduced.

Based on this (limited) stability analysis, one can expect that the
traveling patterns would be found more often when the splay interactions are
taken into account. Furthermore, it can be expected that the traveling
patterns in the parameter region occupied by stationary Turing-like
structures in absence of splay interactions (region $II$ in Fig. \ref
{figdiagfas}a), would be different for weak splay interactions from the
traveling patterns in the parameter region where traveling waves are
observed even without the splay effects (region $I$ in Fig. \ref{figdiagfas}a).
As $\Lambda \rightarrow 0$, traveling waves in the former region $II$
should slow down and become frozen at $\Lambda =0$. Indeed, the analytical
stability investigation shows that, in this region, the velocity of the most
unstable mode is proportional to $\kappa \Lambda \overline{a}$. On the other
hand, the velocity of patterns in the former region $I$ remains finite in
the limit $\Lambda \rightarrow 0$.

Numerical simulations of the model, which represent the main part of the
reported study, agree with the predictions of the stability anaylsis.

\section{Nonequilibrium Phenomena: Numerical Results}
\label{reaction}

To facilitate the comparison with previous simulations of the model without
azimuthal variation\cite{our}, we choose here the same numerical values of
the common parameters $\chi_{2}=0.0052$, $\chi_{0}=2.27$, $\alpha =1.5$ and
$\beta =2$.
The above mentioned changes in the phase diagram due to the inclusion
of the linear splay term indicate that
this parameter region is worth examining with the model presented
here in order to obtain spatio-temporal behaviors involving now composition,
tilt and azimuth modulations.
We examine numerically such parameter region, and the results are summarized
in the three following subsections
according to the value fixed for the splay coupling constant $\Lambda$.
All simulations are perfomed for a system with a linear size $L=5$. 

\subsection{No splay coupling}
\label{nosplay}

When splay coupling is absent ($\Lambda =0$), azimuthal orientation of
elongated molecules is not influenced by variations of the local composition 
$c$ or the local tilt $a$. Indeed, by looking at the second of equations
(\ref{kinfull}) we notice that then all terms on the right hand side of this
equation, except for the elastic term $K\nabla ^{2}\vec{a}$, are
proportional to $\vec{a}$ and therefore cannot change the direction of this
vector. If the azimuthal orientation is initially uniform ($\varphi=const$),
this state is maintained at all times. For such a state, Eqs. (\ref{kinfull})
reduce to the model which was already investigated in Ref.\cite{our} and
new numerical simulations are not needed. If the initial azimuthal
orientation is not uniform ($\varphi\neq const$), the subsequent evolution
of the orientation field is determined only by the elastic term.

\begin{figure}[tbh]
\vspace{-0.05in}
\begin{center}
\epsfxsize = 3.4in
\epsffile{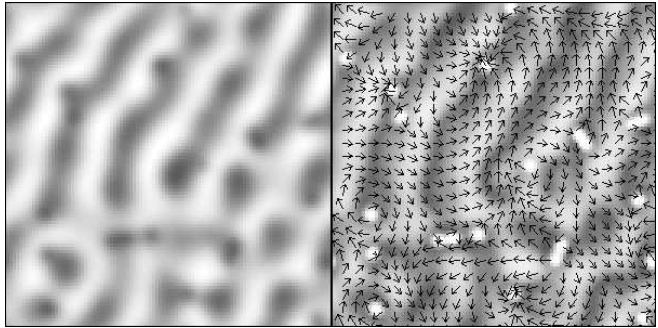}
\end{center}
\caption{A snapshot of concentration and orientation
fields in a pattern of travelling waves that interact with orientational defects;
$\Lambda =0$, $K=0.001$, $\pi _{0}=1.5$ and $\kappa=1 $. The waves propagate from the
upper-left to the lower-right corner of the figure. See also video Fig\ref{figl0}.mpg.}
\label{figl0}
\end{figure}

In numerical simulations, we choose the parameters inside the region $I$
with traveling waves in the phase diagram shown in Fig. \ref{figdiagfas}a, by
fixing $\pi_{0}=1.5$, $\kappa =1$ and taking a small value $K=0.001$ of the
elastic interaction constant. Random distribution of azimuthal orientations
is chosen as the initial condition. The simulation results are presented in
Fig. \ref{figl0} (see also video Fig\ref{figl0}.mpg).
Because of elastic interactions,
the molecules tend to have parallel orientation, and this leads,
after some time, to a pattern
characterized by a number of orientational defects that remains
stationary henceforth. In the center of
a defect, the molecules are non-tilted; the azimuthal direction changes by 
$2\pi$ after passing around a defect. 
The waves, similar to those described in Ref.\cite{our},
travel on the background of the stationary orientational pattern, and their
motion is not generally influenced by the azimuth angle
distribution. The waves, however, break when they pass through a defect.

Inside the region $II$ in the phase diagram in Fig. \ref{figdiagfas}a,
stationary Turing-like patterns representing arrays of droplets are
observed.

\subsection{Weak splay coupling}
\label{moderate}

The most important change in the phase diagram due to the inclusion of the
linear splay term is that the region $II$ occupied by the Turing-like
patterns in absence of splay coupling (Fig. \ref{figdiagfas}a) is transformed
into region $I$ where traveling patterns should be observed
(Fig. \ref{figdiagfas}b). However, as we shall see, the properties of traveling
patterns in the parameter domain are different, for weak coupling, from
those of the traveling waves in the old region $I$.

\begin{figure}[htb]
\vspace{-0.05in}
\begin{center}
\epsfxsize = 3.4in
\epsffile{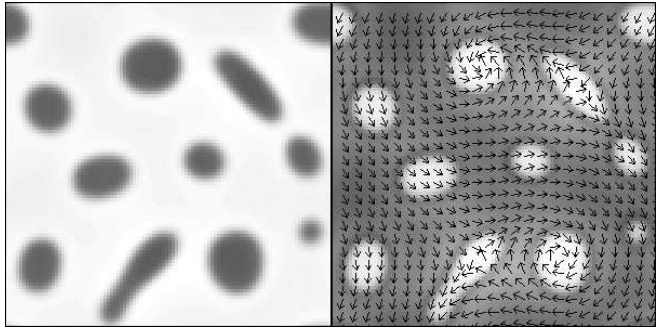}
\end{center}
\caption{A snapshot of concentration and orientation
fields in a pattern of flowing droplets; $\Lambda=K=0.01$, $\pi_0=1.25$ and $\kappa=0.5$.
See also video Fig\ref{figkpetit}.mpg.}
\label{figkpetit}
\end{figure}

Figure \ref{figkpetit} and video Fig\ref{figkpetit}.mpg show the traveling
pattern yielded by numerical
simulation with the parameters $\kappa=0.5$, $\pi_{0}=1.25$, and
$\Lambda=K=0.01$ that correspond to the former region $II$.
The pattern looks like a
flow of droplets, with the local direction of flow determined by the
azimuthal orientation of the elongated molecules in the monolayer. Such
droplet-like structures of large $c$ and small $a$ emerge and move following
the local molecular orientation path. Occasionally, rupture of the droplets,
when they happen to approach divergent points of the molecular orientation
field, is observed. Some of the droplets are pinned by the orientational
defects and exhibit only rotation, but not a translational motion.
Remarkably, the pattern of azimuthal orientation remains frozen under weak
splay coupling, as it was the case in its absence.

In agreement with the linear stability analysis, the velocity of the
emerging droplet structures is an increasing function of the splay coupling
constant $\Lambda$ and the reaction rate constant $\kappa$.
Moreover, we have found that the size of such traveling
structures is affected by the strength of elastic interactions: the larger 
$K$ the bigger are the droplets. For large $K$, however, the uniform state becomes
stable, and no spatial organization is observed.

\begin{figure}[htb]
\vspace{-0.05in}
\begin{center}
\epsfxsize = 3.4in
\epsffile{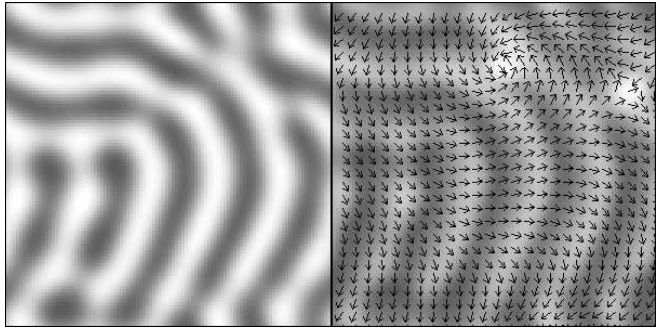}
\end{center}
\caption{A snapshot of concentration and orientation
fields in a pattern of travelling stripes; $\Lambda=K=0.01$, $\pi_0=1.25$ and $\kappa=1$.
See also video Fig\ref{figkgran}.mpg.
}
\label{figkgran}
\end{figure}

Figure \ref{figkgran} and video Fig\ref{figkgran}.mpg show traveling
waves for a larger value of the
reaction rate constant ($\kappa=1$), such that we are now inside the old
region $I$ in Fig. \ref{figdiagfas}a. The morphology of the pattern is now
different and it resembles the pattern of traveling waves in absence of
splay coupling (Fig. \ref{figl0}). The pattern is formed by stripes that move
along the direction determined by local azimuthal orientation of the
elongated molecules. Two orientational defects are seen in the upper right
corner in Fig. \ref{figkgran},and they also correspond to defects of the traveling
stripe pattern of the composition field. The waves are rotating around these
orientational defects. Again, the pattern of azimuthal orientation becomes
frozen after a rapid initial transient and it is not significantly
affected by the traveling waves. Note from the simulation videos
that the stripes move much faster than the droplets in Fig. \ref{figkpetit}.

\begin{figure}[htb]
\vspace{-0.1in}
\begin{center}
\epsfxsize = 2.4in
\epsffile{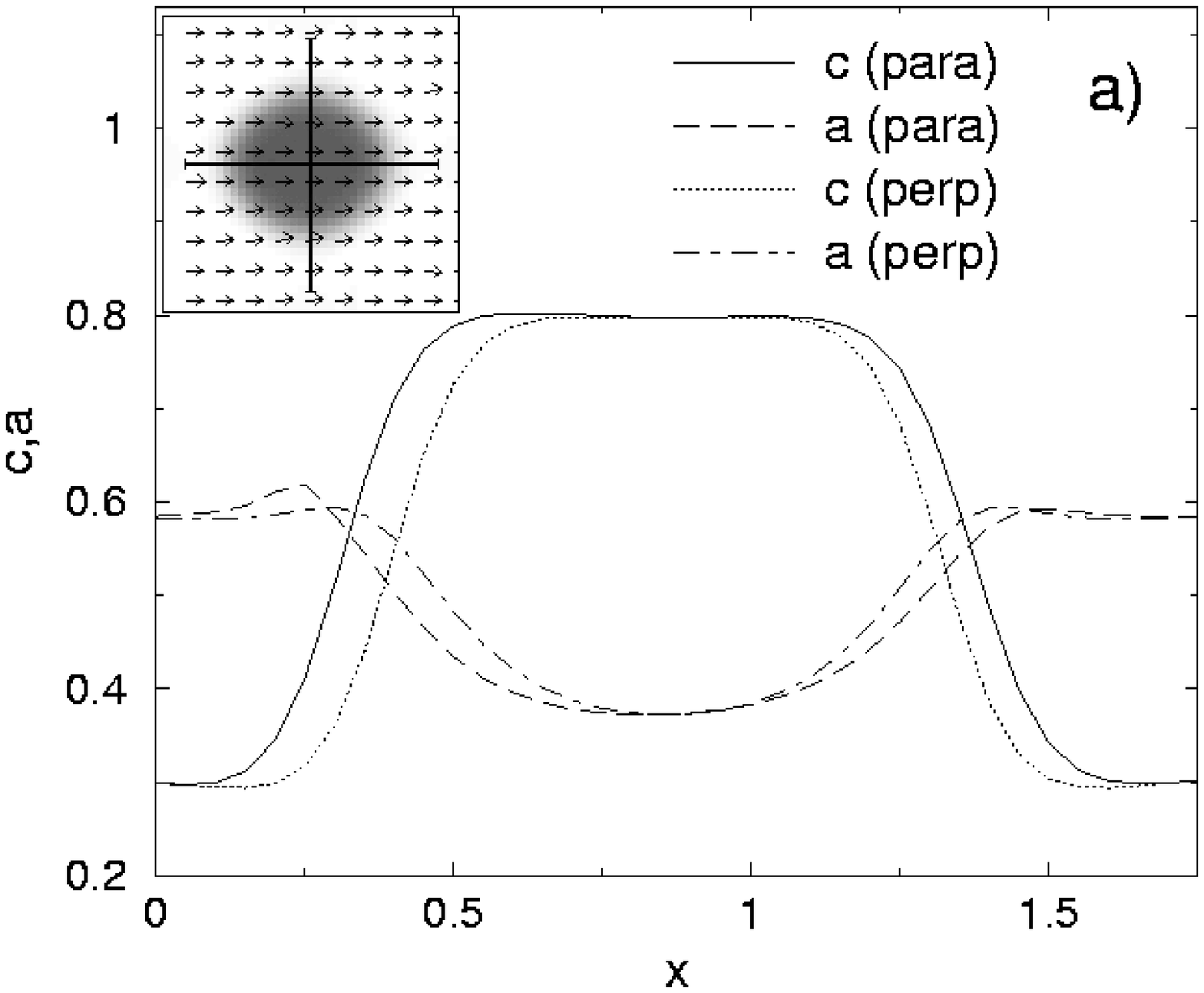}
\epsfxsize = 2.4in
\epsffile{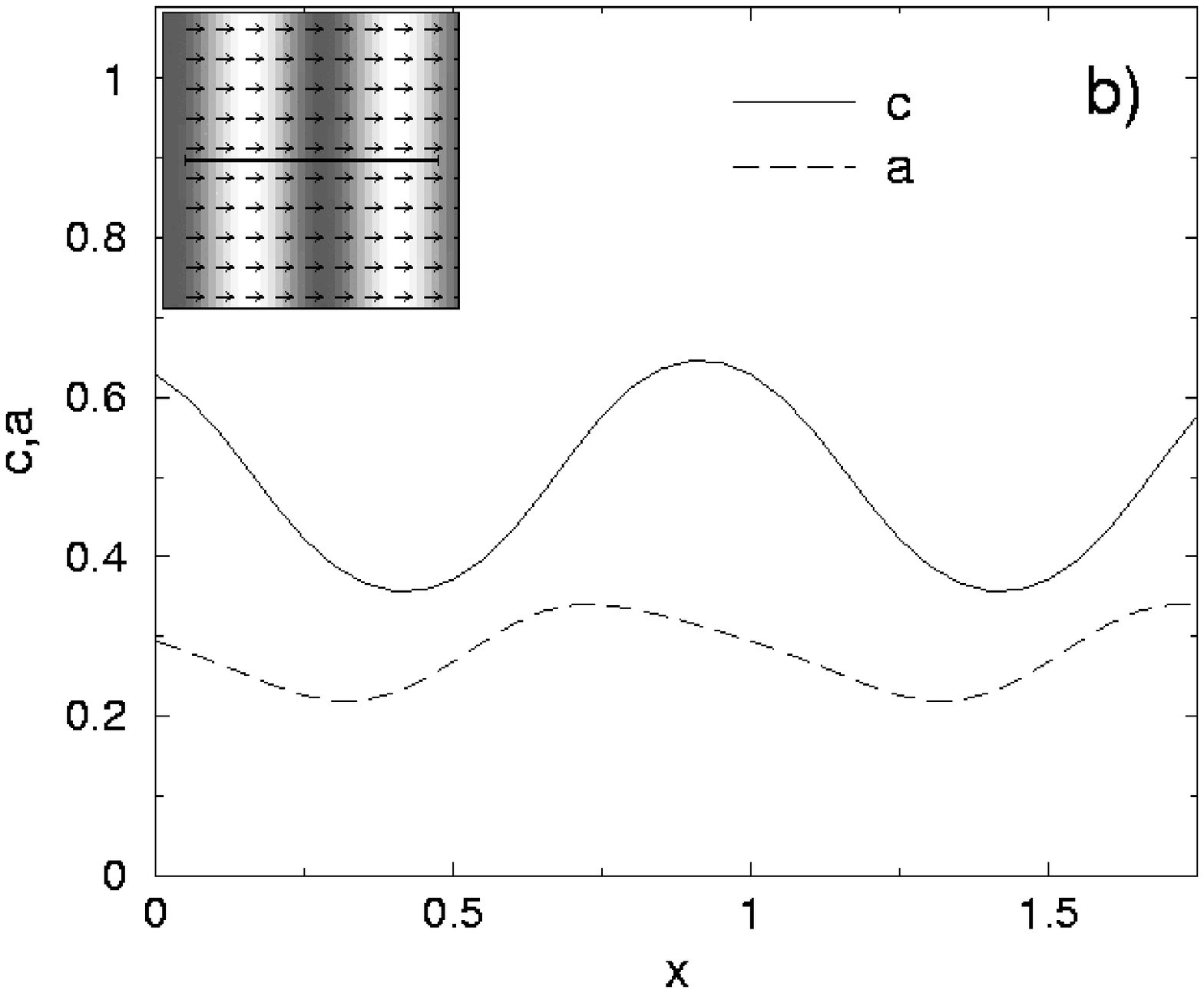}
\end{center}
\caption{Concentration and tilt profiles in the cross sections of (a) flowing droplets
and (b) travelling stripes, shown in Figs. \ref{figkpetit} and \ref{figkgran}.
The respective patterns and the orientations of lines used to make the cross
sections are displayed in the insets. For the droplet, both the profiles along
the lines parallel (``para'') and perpendicular (``perp'') to the motion direction are presented.
The patterns move from left to right.}
\label{figperf}
\end{figure}

Next, we examine more closely the profiles of the composition $c$ and the
tilt $a$ in different traveling patterns. Figure \ref{figperf}a shows such
profiles for a single traveling droplet from Fig. \ref{figkpetit},
also displayed in the
inset in the left upper corner of this figure. The profiles in the cross
sections which are parallel and perpendicular to the motion direction are
presented here. The droplet corresponds to a local increase in the
concentration of elongated molecules $A$ and a local decrease in the tilt of
such molecules. It can be noticed that the tilt is also slightly increased
along a ring surrounding the droplet, and that the droplet is not axially symmetric.
Comparing the profiles for the perpendicular and parallel cross sections in
Fig. \ref{figperf}a, we find that the droplet is slightly elongated in the
direction parallel to the azimuth molecular orientation and, moreover, the
tilt peak in the rear part of the moving droplet is higher than that at
the front. This asymmetry determines the propagation direction of the
droplet. Furthermore, we have checked that the asymmetry gets stronger when the splay
coupling coefficient $\Lambda $ is increased.

The profiles of a traveling stripe from Fig. \ref{figkgran} are displayed
in Fig. \ref{figperf}b.
We see that the profiles exhibit more smooth variation in this
case, and the variation of $c$ and $a$ is closer to harmonical. Note that the
spatial profile of the tilt $a$ is retarded with respect to that of the
local composition $c$ and this again determines the motion direction.


\begin{figure}[htb]
\vspace{-0.05in}
\begin{center}
\epsfxsize = 3.4in
\epsffile{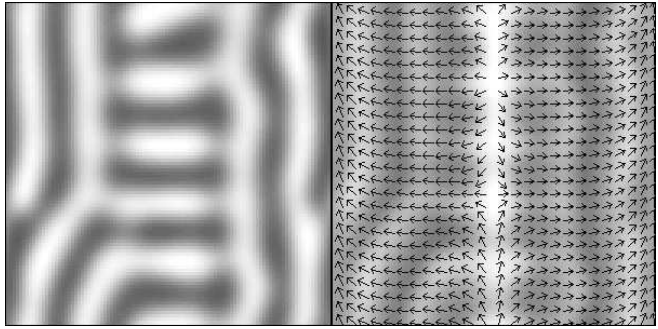}
\end{center}
\caption{A snapshot of concentration and orientation
fields in a pattern representing a linear wave source; $\lambda=0.01$,
$K=0.01$, $\kappa=1$ and $\pi_0=1.25$.
To obtain this pattern special initial conditions
($\varphi=\pi$ for $x<L/2$ and $\varphi=0$ for $x>L/2$) were taken.
The wave structures are generated at the central vertical line and propagate to the side.
See also the video Fig\ref{figvert}.mpg.}
\label{figvert}
\end{figure}

In the simulations described above random initial conditions were chosen.
We have also studied some special initial conditions for the orientation
distribution that give rise to domain sinks or sources. In the simulation
shown in Fig. \ref{figvert} and video Fig\ref{figvert}.mpg,
the initial condition with two
orientational domains ($\varphi=0$ for $x<L/2$ and $\varphi=\pi$ for
$x>L/2$) has been chosen. The boundary between the domains plays here a role
of source emitting waves. For other initial conditions, corresponding to
outward vortices of molecular orientation, we have observed droplets that
are generated in the center of a vortex and travel in the radial direction
out of it. Generally, linear and point defects with positive (negative)
splay $(\vec{\nabla}\cdot \vec{a})$ act as sources (sinks) for the traveling
structures that involve tilt and composition variations.

\subsection{Strong splay coupling}
\label{strong}

\begin{figure}[tbh]
\vspace{-0.05in}
\begin{center}
\epsfxsize = 3.4in
\epsffile{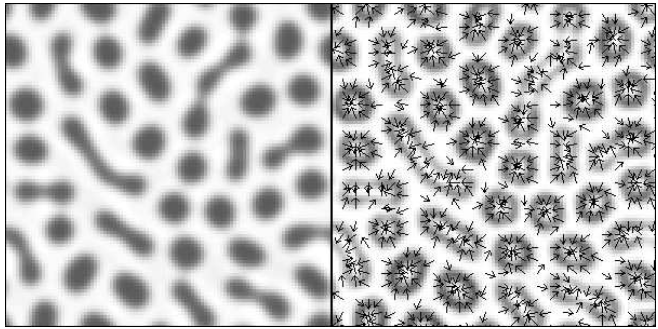}
\end{center}
\vspace{-0.05in}
\caption{A snapshot of concentration and orientation
fields in a stationary array of splay defects;
$\Lambda =1$, $K=0.1$, $\kappa =1$ and $\pi_{0}=1.25$.}
\label{figlgtk}
\end{figure}

When the coefficient $\Lambda$, specifying the intensity of splay coupling,
is further increased, the azimuthal orientation of molecules becomes
influenced by the traveling or stationary patterns. For strong splay
coupling, stationary nonequilibrium patterns are usually observed. An
example of such a pattern is shown in Fig. \ref{figlgtk}. Starting from
random initial conditions, the system first develops a pattern of traveling
stripes. Subsequently, the stripes undergo breakdown and a frozen array of
orientation defects is produced. Inside each defect, the concentration of
the elongated molecules is increased, and in general, these molecules
are oriented towards the center of a defect. This stationary pattern is
favoured in the case of
strong coupling, since it minimizes the splay contribution towards the
free energy of the system.

To avoid the formation of immobile splay defects, a region in the parameter
space, characterized by the wave instability but lying closer to the
boundary of a transition to the nontilted state, can be considered. Because
the elongated molecules are only slightly tilted in this region, their
reorientation is energetically easier than far from the boundary 
$\overline{a}=0$.
If a relatively large value of the reaction rate constant $\kappa$ is
additionally chosen, so that the system is driven further away from thermal
equilibrium, traveling waves that are accompanied by azimuthal reorientation
of molecules are formed.

\begin{figure}[tbh]
\vspace{-0.05in}
\begin{center}
\epsfxsize = 3.4in
\epsffile{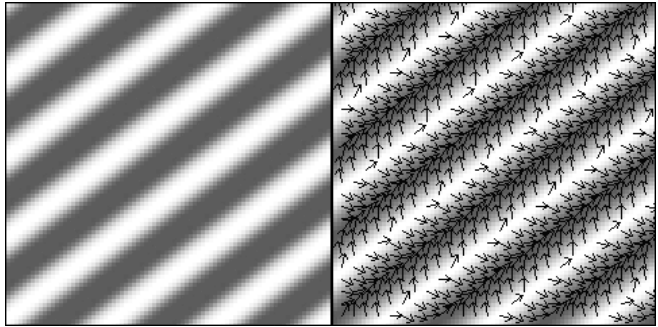}
\end{center}
\caption{A snapshot of concentration and orientation
fields in the patterns of travelling stripes that involves azimuth reorientation;
$\Lambda=1.5$, $K=0.1$, $\kappa=3.5$ and
$\pi_{0}=3.7$. See also the video Fig\ref{figcorner}.mpg.}
\label{figcorner}
\end{figure}

\begin{figure}[tbh]
\vspace{-0.05in}
\begin{center}
\epsfxsize = 2.5in
\epsffile{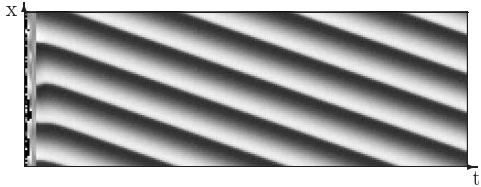}
\end{center}
\caption{Space-time diagram showing the temporal variation of the azimuth
angle along the central vertical cross section in Fig. \ref{figcorner}.
Here, the grey-scale coding is such that $\varphi =0$ is shown as white
and $\varphi =\pi $ is shown as black. The azimuth
angle is defined in such a way that $\varphi =0$ corresponds to
the propagation direction of the stripes.}
\label{figcornerevol}
\end{figure}

An example of the developing wave pattern is shown in Fig. \ref{figcorner}
and video Fig\ref{figcorner}.mpg.
The temporal evolution of this pattern along the central vertical cross
section is additionally displayed in Fig. \ref{figcornerevol}. After a short
transient starting from random initial conditions, the system soon developes
a regular pattern of stripes traveling at a constant velocity. 

\begin{figure}[tbh]
\vspace{-0.1in}
\begin{center}
\epsfxsize = 3.in
\epsffile{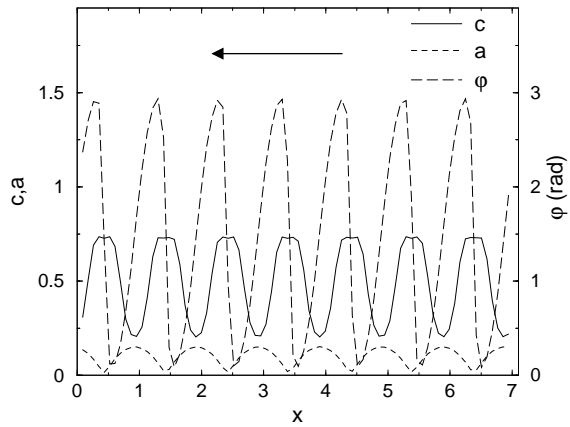}
\end{center}
\caption{Composition, tilt and azimuth profiles for a cross section
perpendicular to the traveling stripes in Fig. \ref{figcorner}. The arrow
indicates the direction of motion of the traveling structures.}
\label{figperfcorner}
\end{figure}

The profiles of the composition $c$, the tilt $a$ and the azimuth angle
$\varphi$ along the line orthogonal to the propagation direction are shown
in Fig. \ref{figperfcorner}. An important difference with the profiles of
equilibrium stationary stripes is that, for traveling stripes, the azimuth
angle does not undergo $2\pi$ rotations from one stripe to another (seen in
Fig. \ref{figperfstripes}). Instead, a periodic angular modulation within the
interval from zero to $\pi$ is observed here. Moreover, it can be noticed
that the composition $c$ and the tilt $a$ vary completely in phase for such
traveling stripes, i.e. the maximum of the concentration field coincides
with a minimum of the tilt distribution. This differs from the behaviour of
such variables in traveling stripes at weak splay coupling (cf. Fig. \ref
{figperf}b). The propagation direction is now fixed by a small shift
observed in the azimuth angle variation with respect to the profiles of
variables $c$ and $a$.


\section{Conclusions}
\label{concl}

We have formulated a theoretical model to describe pattern
formation in illuminated two-component Langmuir monolayers. In this model, 
nonequilibrium pattern formation results from an interplay between phase
separation, optically induced transitions between the two immiscible
conformations, and coupling of phase separation to orientational ordering of
molecules in the monolayer. In comparison with the previous model 
\cite{our}, the current description is more complete because variations of the
azimuthal variation of the elongated molecules in the monolayer are allowed.
We have also incorporated into the model the splay coupling between
azimuthal orientation and variations of local concentration. The situation
corresponding to illumination with non-polarized light has been considered. 

In contrast to the earlier simple model \cite{our}, the investigated system
is characterized, in absence of illumination, by the formation of
equilibrium stationary patterns representing arrays of orientational defects (vortices),
stripes or soliton-like structures with sharp domain walls.
All these patterns are caused by splay interactions and
have a purely energetic origin, so that they can be also interpreted
by considering minima of free energy (cf. \cite{tabe2a,tabe2b,selinger}).

Linear stability investigation of the uniform state of the model under
illumination conditions has been performed. It reveals that, generally,
splay interaction favour the appearance of traveling patterns. Such patterns
are observed even in the parameter regions occupied by stationary structures
in the earlier model \cite{our}. 

Numerical simulations of the proposed model have been undertaken. They show
a rich spectrum of spatio-temporal structures. For weak splay coupling and
relatively low reaction rates, a pattern of droplets slowly flowing along
the local directions determined by azimuthal orientation of molecules has
been observed. If the reaction is stronger, the droplets are replaced by a
pattern of traveling stripes following the local azimuthal orientation.
Interactions between droplets and stripes and the orientational defects have
been seen. It is found that linear and point orientational defects
play a role of sources or sinks of traveling structures in this system.

When splay coupling is weak, the spatial pattern of azimuthal orientation is
not affected by local concentration variations. The temporal evolution of
the orientation pattern in this case is governed by the elastic interactions
between molecules. Starting from a random initial distribution, they produce
after a short transient a stationary pattern of azimuthal orientation that
remains frozen afterwards. In contrast to this, strong splay coupling with
the concentration field leads to the appearance of stationary and
spatiotemporal patterns where variations of azimuthal orientation are
correlated with the the changes in the local composition of the
monolayer. Often, stationary arrays of splay defects are observed in this case.
However, when the
system is close to the orientational phase transition (from the tilted to
the non-tilted phase), patterns of traveling waves accompanied by strong
variation of azimuthal orientation are found in the model. 

This investigation, as well as the earlier Letter \cite{our}, are motivated
by the experimental discovery of traveling structures in illuminated
Langmuir monolayers by Tabe and Yokoyama \cite{tabe1,tabe3}. The
extended model, which we have now explored, is able to reproduce some
essential features of the experiments. In absence of illumination,
equilibrium orientation patterns are yielded by the model. Under
illumination for sufficiently strong splay coupling, traveling waves
accompanied by changes in azimuthal orientation of molecules are also found. The
experiments \cite{tabe1} were performed by using polarized light, and light
polarization significantly influenced the properties of the observed
patterns. In contrast to this, our model does not include the effects of
light polarization and thus corresponds to illumination with non-polarized
light. Though the inclusion of anisotropy effects due to light polarization
is straightforward, the resulting model is more complicated and not yet
discussed in the present paper. 

Langmuir monolayers represent a classical example of soft matter and are
closely related to biomembranes, playing a fundamental role in cell biology.
Therefore, investigations of nonequilibrium pattern formation in such
monolayers under reactive conditions can help in understanding of general
mechanisms of nonequilibrium self-organization in soft matter. Our study
provides evidence that traveling orientational wave patterns may represent a
generic property of Langmuir monolayers that are subjected to chemical
reaction and composition changes. We have found that various
patterns, representing traveling droplets or stripes, wave sources and
sinks, and orientational defects interacting with the traveling structures,
are possible in such systems. 

\section*{Acknowledgments}

This work was supported by the Direcci\'{o}n General de Investigaci\'{o}n
(Spain) through Project BXX2000-0638, and by the Comissionat per
Universitats i Recerca (Generalitat de Catalunya) through Project
1999SGR00041. The study was also supported through the ESF REACTOR network
programme.



\begin{references}

\bibitem{and1} M. Seul and D. Andelman, Science {\bf 267}, 476 (1995).
\bibitem{mikhsc} A. S. Mikhailov and G. Ertl, Science {\bf 272}, 1596 (1996).
\bibitem{glo2} S. C. Glotzer, E. A. Di Marzio and M. Muthukumar, Phys. Rev. Lett. {\bf 74} (11), 2034 (1995).
\bibitem{moto} M. Motoyama and T. Ohta, J. Phys. Soc. Jpn. {\bf 66}, 2715 (1997).
\bibitem{tcong1} Q. Tran-Cong and A. Harada, Phys. Rev. Lett. {\bf 76} (7), 1162 (1996).
\bibitem{dewell} J. Verdasca, P. Borckmans and G. Dewel, Phys. Rev. E {\bf 52}, R4616 (1995).
\bibitem{mexPRE} M. Hildebrand, A. S. Mikhailov and G. Ertl, Phys. Rev. E {\bf 58}, 5483 (1998).
\bibitem{hild} M. Hildebrand, A. S. Mikhailov and G. Ertl, Phys. Rev. Lett. {\bf 81} (12), 2602 (1998).
\bibitem{mikhstat} M. Hildebrand and A. S. Mikhailov, J. Stat. Phys. {\bf 101}, 599 (2000).

\bibitem{okuz} T. Okuzono and T. Ohta, Phys. Rev. E {\bf 64}, 045201(R) (2001).
\bibitem{ohta2} T. Okuzono and T. Ohta, Phys. Rev. E {\bf 67}, 056211 (2003).

\bibitem{kag1} V. M. Kaganer, H. M\"{o}hwald and P. Dutta, Rev. Mod. Phys. {\bf 71}, 779 (1999).
\bibitem{tabe1} Y. Tabe and H. Yokoyama, Langmuir {\bf 11}, 4609 (1995).
\bibitem{tabe2a} Y. Tabe and H. Yokoyama, J. Phys. Soc. Japan {\bf 63}, 2472 (1994).
\bibitem{tabe2b} Y. Tabe, N. Shen, E. Mazur and H. Yokoyama, Phys. Rev. Lett. {\bf 82}, 759 (1999).
\bibitem{tabe3} Y. Tabe, T. Yamamoto and H. Yokoyama, New J. Phys. {\bf 5}, 65 (2003).
\bibitem{our} R. Reigada, F. Sagu\'{e}s and A. S. Mikhailov, Phys. Rev. Lett. {\bf 89}, 038301 (2002).

\bibitem{laidler} K. J. Laidler, {\it Chemical Kinetics},
(3rd Edition, HarperCollins Pub., NY, 1987).
\bibitem{maack2} S. Malkin and E. Fischer, J. Phys. Chem. {\bf 66}, 2482
(1962).
\bibitem{maack} J. Maack, R. C. Ahuja and H. Tachibana, J. Phys. Chem. {\bf 99}, 9210
(1995).

\bibitem{mikh6} M. Hildebrand and A. S. Mikhailov, J. Chem. Phys. {\bf 100}, 19089 (1996).

\bibitem{meyer} R. Meyer and P. Pershan, Solid State Commun. {\bf 13}, 989 (1973).
\bibitem{pets} G. A. Hinshaw, Jr., and R. G. Petschek, Phys. Rev. A {\bf 39}, 5914 (1989).
\bibitem{najjar} R. Najjar and Y. Galerne, Europhys. Lett. {\bf 55}, 355 (2001).
\bibitem{selinger} J. V. Selinger, Z. G. Wang, R. F. Bruinsma and C. M. Knobler,
Phys. Rev. Lett. {\bf 70}, 1139 (1993).

\bibitem{esc3d} R. B. Meyer, Philos. Mag. {\bf 27}, 405 (1973).

\bibitem{proper} J. Ign\'{e}s-Mullol, {\it et al.}, in preparation.

\bibitem{wal} D. Walgraef, {\it Spatio-Temporal Pattern Formation},
(Springer-Verlag, New York, 1997).


\end{references}
\end{document}